# Portosystemic hepatic encephalopathy model shows reversal learning impairment and dysfunction of neural activity in the prefrontal cortex and regions involved in motivated behavior

**M Méndez PhD,* [1] M Méndez-López PhD, [2] L López PhD, [1] MA Aller PhD, [3] J Arias PhD, [3] JL Arias PhD [1]**

[1] Laboratorio de Neurociencias. Departamento de Psicología. Universidad de Oviedo. Plaza Feijoo s/n. 33003 Oviedo, Spain
[2] Departamento de Psicología y Sociología. Universidad de Zaragoza. Campus Ciudad Escolar s/n. 44003 Teruel, Spain
[3] Departamento de Cirugía I, Facultad de Medicina, Universidad Complutense de Madrid. Ciudad Universitaria s/n. 28040 Madrid, Spain

---

***Corresponding autor:** Marta Méndez, PhD
Laboratorio de Neurociencias. Departamento de Psicología
Plaza Feijoo s/n 33003 Oviedo, Spain
Telephone: (+34) 985103212; Fax (+34) 985104144
Email: mendezlmarta@uniovi.es

---

## Abstract

Hepatic encephalopathy (HE) is a neurological complication that affects attention and memory. Experimental animal models have been used to study HE, the most frequent being the portacaval shunt (PCS). In order to determine learning impairment and brain functional alterations in this model, we assessed reversal learning and neural metabolic activity in a PCS rat model. PCS and sham-operated rats (SHAM) were tested for reversal learning in the Morris water maze. Brains were processed for cytochrome oxidase (CO) histochemistry. The PCS group presents reversal learning impairment and CO activity reduction in prefrontal cortex, ventral tegmental area and accumbens shell nucleus. These results suggest that this model of portosystemic HE shows learning impairment that could be linked to dysfunction in neural activity in the prefrontal cortex and regions involved in motivated behavior.

## Introduction

Many patients with liver cirrhosis suffer from a neuropsychiatric syndrome called hepatic encephalopathy (HE). As HE progresses, intellectual abilities deteriorate and patients show neuropsychological disturbances that affect attention, memory and visuospatial orientation.(1-3) In spite of advances in recent years, the factor or factors that determine the development of HE are still unclear and it is necessary to recur to animal experimentation in order to clarify the brain substrates of the cognitive deficits found in HE.

Diverse experimental animal models have been used to study both the behavioral and the brain dysfunctions that occur in HE. Perhaps the most frequently used is that of portacaval shunt, considered one of the best models of chronic minimal Type B HE,(4) which corresponds to encephalopathy associated with portosystemic shunt. Although portacaval-shunted rats present difficulties to learn different types of tasks, such as passive and active avoidance or a conditional discrimination task in the Y maze,(5-7) little research has focused on the learning deficits found in the Morris Water Maze (MWM).(8-10)

The goal of this work was to evaluate spatial reversal learning in a rat model of portacaval anastomosis and to study oxidative metabolism of various brain limbic system regions involved in memory process by histochemical labeling of cytochrome oxidase (CO) (11). This technique can be used as an index of regional functional activity in the brain, reflecting changes in metabolic capacity induced by sustained energy requirements of the nervous system associated with spatial memory.(12,13)

## Materials and methods

We used 14 male Wistar rats from the vivarium of Oviedo University that were maintained under standard laboratory conditions (22±2ºC temperature, 65-70% relative humidity, and 12 h light/dark cycle). The procedures were carried out according to the European Community Council Directive 86/609/EEC and were approved by the local committee for animal studies (Oviedo University).

The animals were randomly distributed into two groups: sham-operated (SHAM group, n=6) and end-to-side portacaval shunt (PCS group, n=8). The surgical procedures were carried out as described previously.(14) Behavioral evaluation was performed 45 days after intervention.

Reversal learning was evaluated in the Morris Water Maze (MWM).(15) This maze can be used to assess learning deficit in experimental models of disease.(16) The maze consists of a circular pool (diameter=150 cm) partially filled with water (30 cm deep, 22 ± 1ºC) located in a room with numerous visual cues. The pool was virtually divided into four quadrants (quadrants A, B, C, and D). The animals were given 4 acquisition trials per training session or day, for up to 4 days, to learn the location of the submerged platform hidden in the center of quadrant D. For each trial, the rat was placed in the pool at one of four possible locations, randomly ordered, and then given 60 s to find the platform. If the platform was not found in 60 s, the rat was placed on the platform for15 s. The inter-trial interval lasted 30 s. Latency to find the platform, defined as escape latency, distance covered, and velocity were recorded. Daily, at the end of the session, rats were given a 25-s probe trial in the maze with the platform removed. In this probe trial, selective quadrant search was evaluated by measuring the percentage of time spent in each quadrant of the pool. After this training, the rats were tested for reversal learning using the same procedure. The animals were given 4 acquisition trials. The hidden platform was located in the quadrant opposite to its previous location, quadrant C. As before, rats were given a probe trial that was conducted at the end of the session.

Ninety minutes after the end of the reversal task, the animals were decapitated, brains were removed intact, frozen rapidly in isopentane (Sigma-Aldrich, Germany) and stored at -40ºC. Coronal sections (30 µm) of the brain were cut at -20ºC in a cryostat (Leica CM1900, Germany). We used a quantitative CO histochemistry.(11) As previously described,(17) sections and brain standards (sets of Wistar rat brain tissue homogenate standards that were cut at different thickness: 10, 30, 40 and 60 µm) were fixed with glutaraldehyde and rinsed in phosphate buffer. They were then immersed in Tris buffer solution containing cobalt chloride. Then, they were incubated in a solution containing cytochrome c, catalase and diaminobenzidine thetrahydrochloride (Sigma-Aldrich, Spain) dissolved in phosphate buffer (pH 7.6, 0.1 M). Lastly, the slides were fixed, dehydrated and cover-lipped with Entellan.

Quantification of CO histochemical staining intensity was carried out by densitometric analysis using a computer-controlled image analysis workstation (MCID, InterFocus Imaging Ltd., Linton, England). The mean optical density (OD) of each structure was measured on the right side of the bilateral structures using three consecutive sections of each animal. In each section, 4 readings were taken using a square-shaped sampling window that was adjusted for each region. A total of 12 measurements were taken per region and averaged to obtain a mean. OD values were then converted to CO activity units, determined by the enzymatic activity of the standards, which was measured spectrophotometrically.

The stained coronal sections, from anterior to posterior, corresponded to Bregma levels (18) 3.24 mm (Prelimbic cortex (PL) and Infralimbic cortex (IL)), 2.28 mm (Accumbens Core (ACc) and Accumbens Shell (Acs)), -1.92 mm (Anterodorsal Thalamic nucleus (TAD), Anteroventral Thalamic nucleus (TAV) and Anteromedial Thalamic nuleus (TAM)), -3.84 mm (Hippocampal CA1 (CA1), Hippocampal CA3 (CA3) and Hippocampal Dentate Gyrus (DG)), -4.44 mm (Medial Mammillary nucleus (MMn), Lateral Mammillary nucleus (LM) and Supramammillary nucleus (SuM)), and -4.8 mm (Ventral Tegmental Area (VTA)).

## Results

For the initial hidden platform test, the latencies to reach the hidden platform for each day (average of 4 trials) were analyzed with a two-way repeated measures ANOVA. When a significant session effect was found, a further repeated measures ANOVA was conducted for each group. There were differences in escape latencies between PCS and SHAM [$F(1,12)=10.421$, $P=0.007$]. PCS presented longer escape latencies than SHAM. The variable day also showed a significant effect [$F(3,36)=9.289$, $P<0.001$]. The SHAM group showed an improvement over the days [$F(3,15)=19.782$, $P<0.001$], presenting longer latencies to target on Days 1 and 2 compared to the rest of the days ($P<0.05$). Similarly, the PCS group showed a reduction in latencies over the days [$F(3,21)=4.542$, $P=0.013$]. The PCS group presented shorter latencies on Day 4 compared to Day 1 ($P<0.05$) (**Fig. 1**). In summary, SHAM rats reduced their latencies to target as of Day 3 whereas the PCS group showed a reduction in latencies on Day 4.
The distance covered (mean±SE) by SHAM (1232±127 cm) and PCS (1224±111 cm) was similar [$t(12)=0.049$, $P= 0.961$]. Also, the SHAM (56±5 cm/s) group showed similar swimming velocity to PCS (54±3 cm/s) [$t(12)=0.360$, $P= 0.725$].
The time spent in each of the four quadrants during the probe test was analyzed separately for each group and day using a one-way ANOVA design, and Tukey's test was applied as post hoc test. The PCS group learned the location of the platform on Day 4. On this day, the PCS group spent more time in the quadrant where the platform was located, Quadrant D, compared to the rest of the quadrants [$F(3,28)=16.069$, $P<0.001$]. With respect to the SHAM group, spatial learning was shown as of Day 3 [$F(3,20)=13.825$, $P<0.001$] and Day 4 [$F(3,20)=15.806$, $P<0.001$]. Tukey's test showed that Quadrant D was preferred by the SHAM rats with respect to A, B, and C ($P<0.05$) (**Fig. 2**).
There were no differences between the PCS and SHAM groups in latencies to reach the hidden platform during the acquisition trials of the reversal test [$t(12)=-1.259$, $P=0.232$] (**Fig.3.a**). However, the groups differed in the hidden platform test. The SHAM rats presented differences in the time of permanence in the quadrants of the pool [$F(3,20)=12.347$, $P<0.001$]. They preferred the new platform position or Quadrant C versus the remaining Quadrants A, B, and D ($P<0.008$). However, the PCS group did not learn the new location of the platform. The PCS group presented differences in the time of permanence in the quadrants of the pool [$F(3,28)=20.538$, $P<0.001$]. The PCS group preferred Quadrant C and D equally, as both quadrants differed from A and B ($P<0.05$) (**Fig.3.b**).
CO activity results were analyzed by a Student *t*-test. The PCS group showed lower CO activity compared to the SHAM group in the PL [$t(12)=-3.172$, $P= 0.008$] and IL cortex [$t(12)=-2.464$, $P= 0.030$], ACs nucleus [$t(12)=-2.220$, $P=0.046$] and VTA [$t(12)=-2.465$, $P= 0.03$] (Fig. 4). No differences were found between groups in the remaining regions studied (**see Table 1**).

## Discussion

Our work revealed the presence of cognitive alterations in the processes of reversal learning in the MWM in a model of Type B hepatic encephalopathy by portacaval anastomosis. This learning impairment is accompanied by decreased CO activity of the medial prefrontal cortex, accumbens shell nucleus, and ventral tegmental area.
The PCS rats were unable to remember the new location of platform during the probe test of the reversal task. These differences between the PCS and SHAM groups cannot be justified by motor problems, as the groups did not differ in their velocity or the distance covered in the MWM. Therefore, although motor problems are among the neurological alterations characteristic of portacaval anastomosis,(19) learning impairment was not due to motor activity alterations in the experimental model used, as shown in another study.(8)
The PCS rats presented learning delay compared to the SHAM rats. We could observe that the PCS group was able to remember the location of the hidden platform on the final day of the task, which reveals a delay in spatial learning by allocentric information, supporting previous results.(8) However, studies aimed at evaluating spatial learning of PCS in the MWM present contradictory results. Some studies report an absence of differences between the SHAM and PCS groups,(20) whereas other studies reveal memory deficits.(9) Perhaps the disparity among data is due to differences in the MWM procedure (duration of intertrial intervals, number of training trials, exploration time evaluated, or variables measured).

Once spatial learning was established, we moved the hidden platform to the opposite quadrant of the pool in the reversal task. Thus, we could observe that, in contrast to the SHAM group, which rapidly learned the new location, the PCS group showed preference for the previous platform position. The PCS model of HE shows perseveration in the previously reinforced quadrant, confirming that spatial reversal is affected. To our knowledge, no studies to date have assessed reversal learning in a model of HE by PCS. The PCS model presents problems in the ability to adjust to changing environment and absence of cognitive flexibility, displaying a reduced ability to relearn the new spatial position.

Regarding animal studies, a deficit of cognitive flexibility has been found in a model of toxic liver failure by thioacetamide treatment.(21) Some authors argue that memory disturbances are not a major symptom of HE in humans,(3) whereas other authors state that patients with cirrhosis present poorer performance in several memory tests.(1,22)

One study has connected memory impairment in an HE model by thioacetamide administration with oxidative metabolism.(17) Other studies have reported brain oxidative metabolism impairment associated with behavioral alterations in PCS rats.(23,24) In the present work, we showed that CO activity was lower in medial prefrontal cortex, accumbens region, and ventral tegmental area in the PCS model compared to SHAM after performing the spatial reversal task in the MWM. This might be due to an inefficient acquisition of the reversal task by the PCS group. These results could be explained by alterations in the medial prefrontal cortex, because one of the main manifestations of prefrontal deficits is perseveration, expressed as the loss of cognitive flexibility associated with the incapacity to inhibit previously learned responses.(25) It has been widely demonstrated that the medial prefrontal cortex, prelimbic, and infralimbic regions are involved in spatial reversal.(26,27)

The ventral tegmental area, part of the dopaminergic system and connected to the prefrontal cortex, mediates motivated behavior and is involved in spatial working memory performance.(28) The accumbens is considered an interface between the limbic and the motor system and plays a role in motivated and goal-directed behavior, being the main neuroanatomical substrate of adaptive responding.(29,30) Inputs from the prefrontal cortex and ventral tegmental area to the accumbens nuclei are essential for the transformation of a prospective plan of action into appropriate behavioral output. This was demonstrated in working memory tasks that require flexible use of information.(31) Therefore, we could suggest that PCS rats present impairment of a brain network responsible for adaptation to changing circumstances, a necessary process to solve reversal tasks.

Related to this, it is well known that PCS rats present a rise in cerebral ammonia that impairs induction of NMDA receptor-dependent long-term potentiation in the hippocampus and also alters the neural glutamate-nitric oxide cyclic GMP pathway, a pathway involved in learning and memory.(6) In addition, hyperammonemia produces alteration of extracellular concentration of neurotransmitters by metabotropic glutamate receptors in the nucleus accumbens and impaired motor control by this nucleus.(32) As demonstrated, PCS rats present disruption of the normal circuit that connects nucleus accumbens and prefrontal cortex.(5)

In the reversal task, the animals continued to respond to the previously reinforced location, showing a deficit in cognitive flexibility represented by perseverative behavior. As this capacity is clearly associated with the medial prefrontal cortex (mainly the prelimbic and infralimbic cortex), it is feasible to consider a possible dysfunction of the prefrontal cortex in hepatic insufficiency. In this sense, these results are in accordance with the study of Peixoto et al.,(33) which confirmed prefrontal dysfunction in subjects with hepatic cirrhosis in tasks involving cognitive flexibility, and with the work of Zhang et al. (34) using a modified Stroop task to assess capacity of inhibition. In conclusion, the results obtained show that the model of Type B HE by portacaval anastomosis presents alterations in spatial reversal and dysfunction of the neural activity involved in motivated behavior.

**Acknowledgements**

This research was supported by current Spanish Ministry of Science and Innovation and FEDER (SEJ2007-63506 and PSI2010-19348) and FMMA (AP/6977-2009).

## Figures and Tables

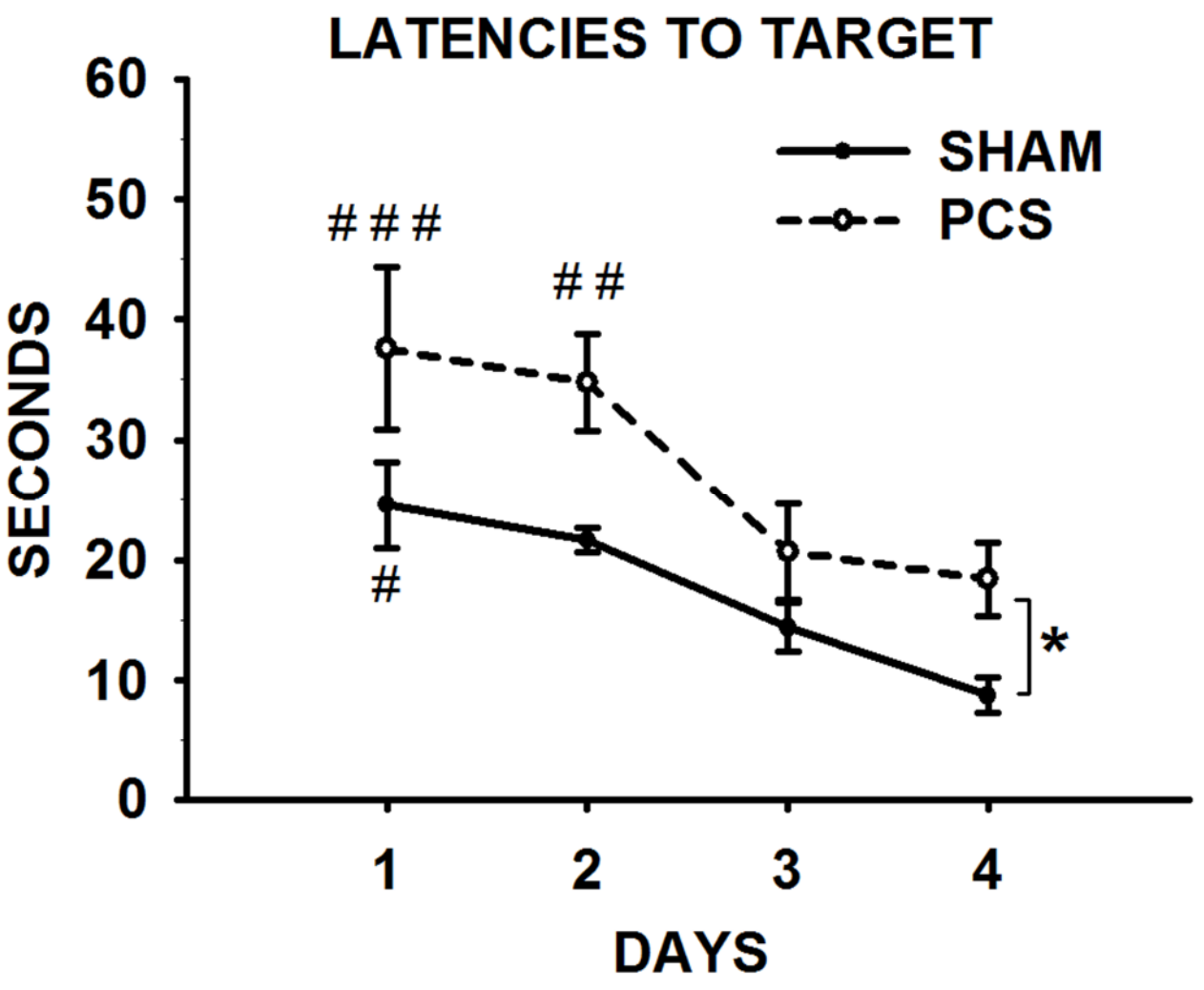

**Figure 1.** Escape latencies during the initial hidden platform test that was carried out during four training sessions or days. Significance of differences between PCS and SHAM (*p<0.05) in escape latencies (mean ± SEM) during the four training sessions of the reference memory task. Significance of differences between sessions are shown (# p<0.05). SHAM rats reduce their latencies to target as of Day 3 whereas the PCS group showed a reduction in latencies on Day 4.

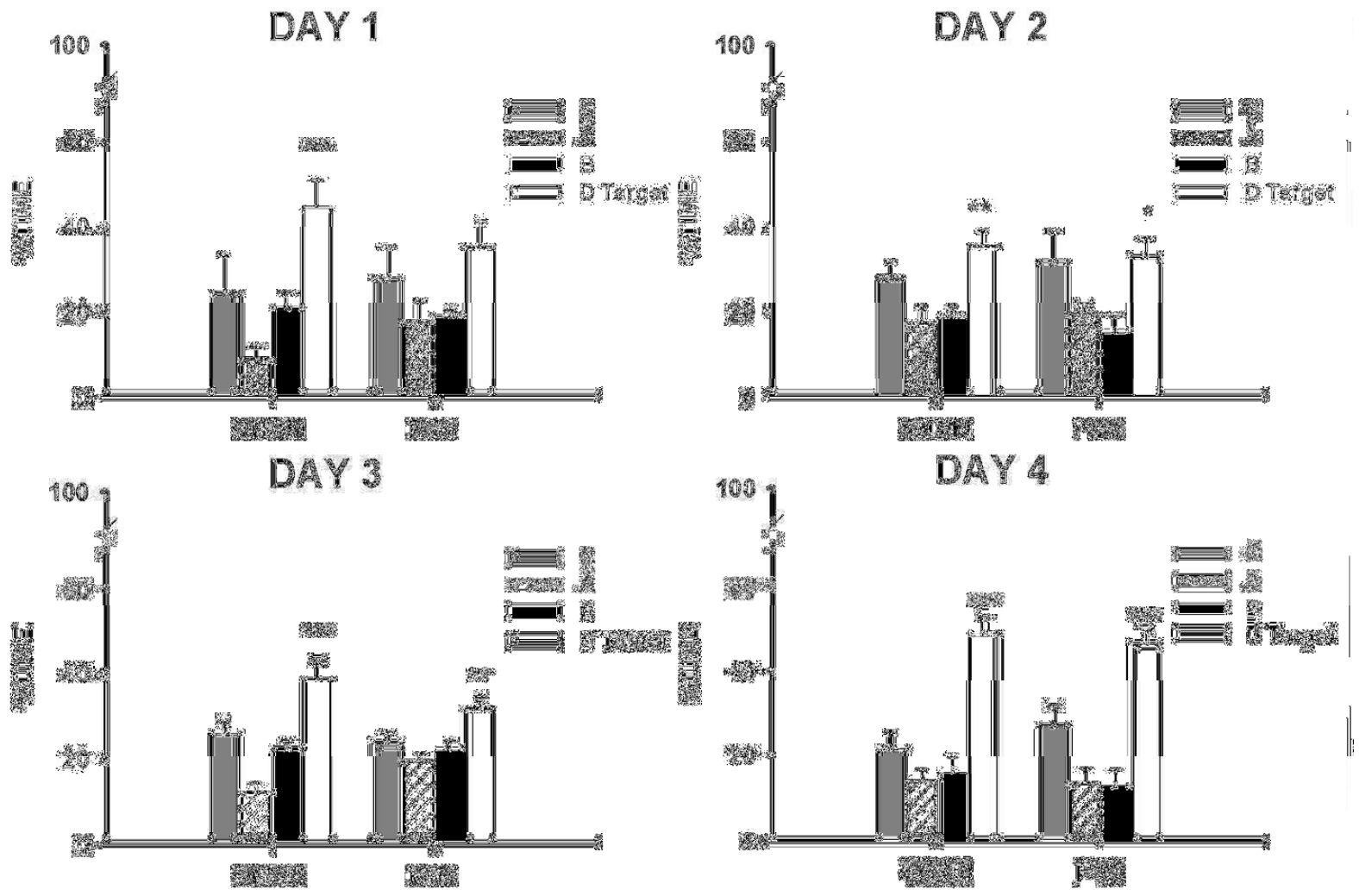

**Figure 2.** Probe tests in the spatial reference memory test during four training sessions (mean ± SEM). In the probe tests, three asterisks above the bar indicate a statistically significant difference in the time spent in Quadrant D (target) compared to time spent in any of the other quadrants (***p<0.05). Two asterisks above the bar indicate a statistically significant difference in the time spent in Quadrant D compared to the time spent in two of the other three quadrants (**p<0.05). One asterisk above the bar indicates significant difference in the time spent in Quadrant D compared to the time spent in one of the other three quadrants (*p<0.05).

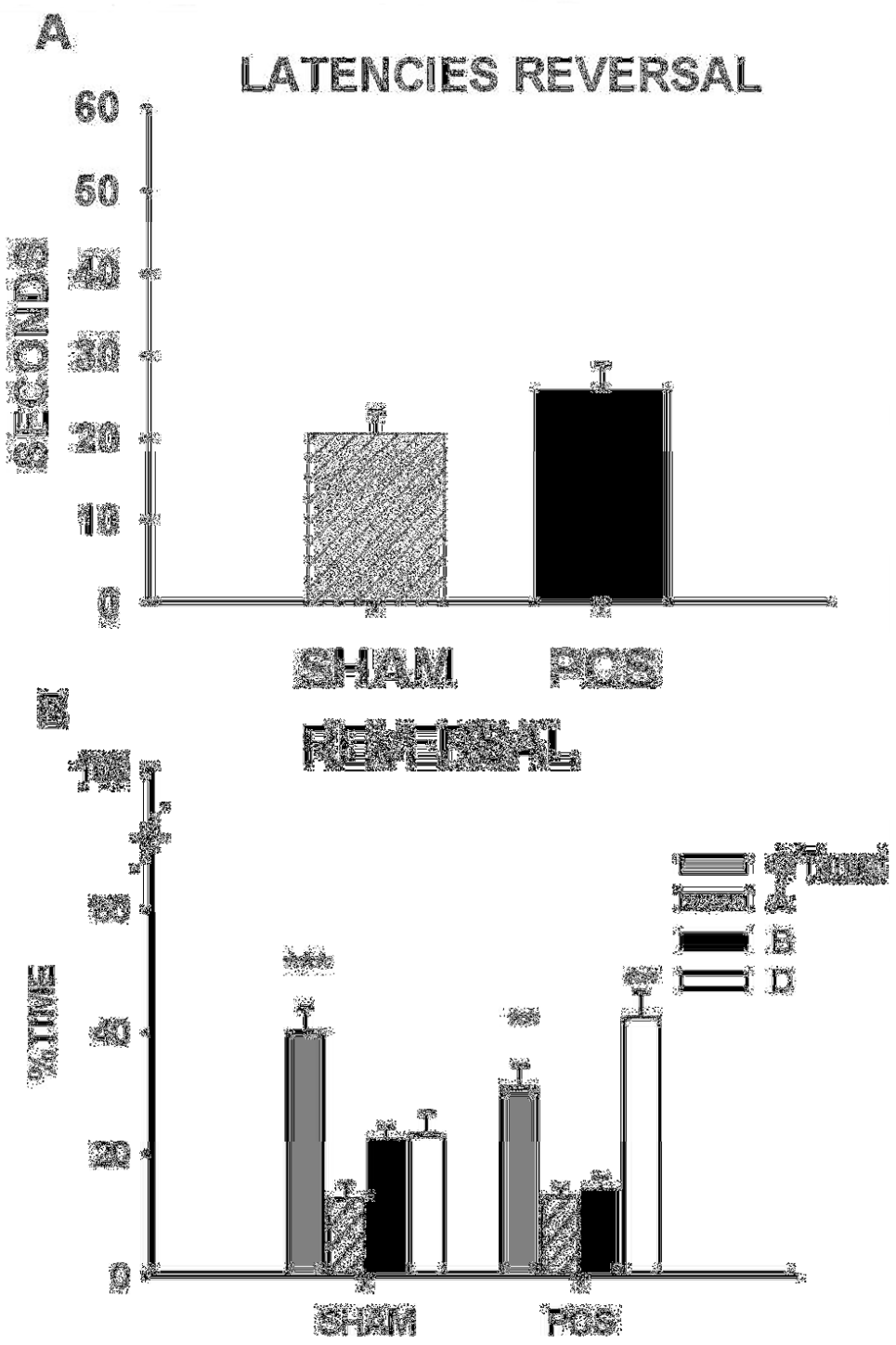

**Figure 3.** (A) Escape latencies (mean ± SEM) of PCS and SHAM during the acquisition trials of the reversal test. (B) Probe test in the spatial reversal task (mean ± SEM). Three asterisks above the bars indicate a statistically significant difference in the time spent in Quadrant C (Target) compared to time spent in any of the other three quadrants (*** p<0.05). Two asterisks above the bar indicate a significant difference in the time spent in Quadrants C or D (previous target) compared to time spent in two of the other three quadrants (**p<0.05).

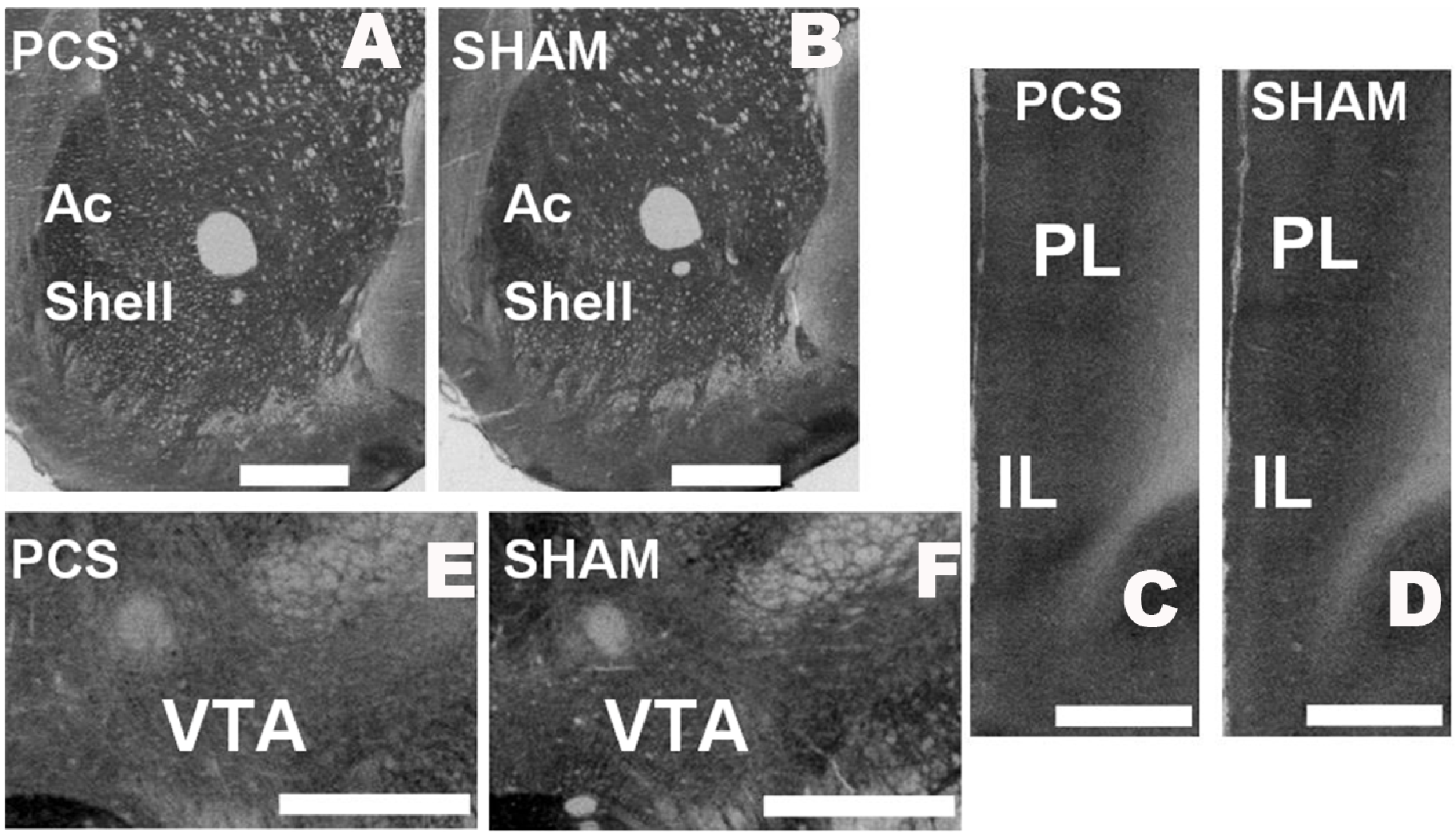

**Figure 4.** Cytochrome oxidase (CO) histochemistry in portacaval shunt (PCS) and sham-operated (SHAM) rats. CO activity in Accumbens Shell (Ac Shell) (A, B), Prelimbic (PL) and Infralimbic (IL) cortex (C, D) and Ventral Tegemental Area (VTA) (E, F). Scale bar: 1 mm.

Table 1. Regional CO activity in the two groups expressed by mean ± SEM.

| | **PCS** | **SHAM** | |
|---|---|---|---|
| Regions | Mean ± SEM | Mean ± SEM | Student *t*-test value; *p* value |
| Prelimbic cortex | 13.719±0.376 | 15.980±0.657 | -3.172; 0.008 * |
| Infralimbic cortex | 12.756±0.402 | 14.359±0.527 | -2.464; 0.030 * |
| Accumbens Core | 17.446±0.532 | 18.665±0.886 | -1.245; 0.237 |
| Accumbens Shell | 21.770±0.553 | 23.670±0.657 | -2.220; 0.046 * |
| Anterodorsal Thalamic nucleus | 33.069±1.056 | 32.655±1.794 | 0.210; 0.837 |
| Anteroventral Thalamic nucleus | 24.903±1.369 | 26.822±2.015 | -0.817; 0.430 |
| Anteromedial Thalamic nucleus | 14.781±0.345 | 15.860±2.002 | -0.614; 0.550 |
| Hippocampal CA1 | 13.737±0.605 | 14.852±0.624 | -1.260; 0.232 |
| Hippocampal CA3 | 11.579±0.573 | 11.687±0.445 | -0.141; 0.890 |
| Hippocampal DG | 20.679±0.696 | 20.864±0.677 | -0.186; 0.855 |
| Medial Mammillary nucleus | 28.837±1.155 | 25.399±1.169 | 2.051; 0.063 |
| Lateral Mammillary nucleus | 30.925±1.259 | 27.175±1.908 | 1.709; 0.113 |
| Supramammillary nucleus | 13.254±0.509 | 14.888±1.203 | -1.377; 0.194 |
| Ventral Tegmental area | 7.797±0.624 | 10.241±0.788 | -2.465; 0.030 * |

Student *t*-test values and *p* values are shown. Significant differences between the groups are represented (*).